\documentclass[useAMS,usenatbib]{mn2e}
\usepackage{textcomp,enumitem}
\include{journal_abbv}
\usepackage{graphicx,natbib}	% Including figure files
\usepackage{amsmath}	% Advanced maths commands
\usepackage{amssymb}	% Extra maths symbols
\usepackage{multicol}
\usepackage{multirow}% Multi-column entries in tables
\usepackage{bm}		% Bold maths symbols, including upright Greek
\usepackage{pdflscape}	% Landscape pages
\usepackage{float}
\usepackage[colorlinks=true,citecolor=blue]{hyperref}
\usepackage[caption = false]{subfig}
\usepackage{longtable}
\usepackage{xcolor}
\usepackage{ulem}

\vbadness=\maxdimen

\newcommand{\apj}{ApJ}
\newcommand{\apjs}{ApJS}
\newcommand{\apjl}{ApJL}
\newcommand{\aap}{A{\&}A}

\newcommand{\mnras}{MNRAS}
\newcommand{\aj}{AJ}
\newcommand{\araa}{ARA\&A}
\newcommand{\pasp}{PASP}

\newcommand{\nat}{Nature}

\newcommand{\pasa}{PASA}
\newcommand{\apss}{apss}

\title[Multi-wavelength Study of Mrk 212]
 {A Multi-wavelength Study of the Dual Nuclei in Mrk 212}
\author[Rubinur et al.]{{K. Rubinur}$^{1,2,3}$\thanks{E-mail: rubinur@ncra.tifr.res.in}, {P. Kharb}$^3$, {M. Das}$^1$, {P.T Rahna}$^{1,6}$, {M. Honey}$^4$, {A. Paswan}$^5$, \newauthor{S. Vaddi}$^3$, {J. Murthy}$^1$\\
$^1$ Indian Institute of Astrophysics, Koramangala II Block, Bangalore 560034, India \\
$^2$ Pondicherry University, R. Venkataraman Nagar, Kalapet, Pondicherry 605014, India \\
$^3$ National Centre for Radio Astrophysics - Tata Institute of Fundamental Research (NCRA-TIFR), S. P. Pune University Campus, \\Ganeshkhind, Pune 411007, India\\
$^4$Syed Abdul Rahiman BafakhyThangal Memorial Government College Koyilandy, P.O.Muchukunnu, Koyilandy, Kozhikode 673307, India\\
$^5$Inter-University Centre for Astronomy and Astrophysics, Ganeshkhind, Post Bag 4, Pune 411007, India\\
$^6$CAS Key Laboratory for Research in Galaxies and Cosmology, Shanghai Astronomical Observatory, 80 Nandan Road, Shanghai 200030, China}

\begin{document}
\label{firstpage}
\maketitle
\begin{abstract}
We present radio observations of the galaxy merger remnant Mrk~212 with the Karl G. Jansky Very Large Array (VLA) and the upgraded Giant Meter Radio Telescope (uGMRT). Mrk~212 has two previously known radio sources associated with the two optical nuclei, S1 and S2, with a projected separation of $\sim$6 kpc, making it a dual active galactic nuclei (AGN) candidate. Our new 15 GHz VLA observations reveal that S1 is a double radio source centred around the optical nucleus; its total extent is $\sim$750~parsec and its average 1.4$-$8.5 GHz spectral index is $-0.81\pm0.06$. S1 therefore, resembles a compact symmetric object (CSO). The 15~GHz VLA image identifies the radio source at S2 to be a compact core. Our radio observations therefore strongly support the presence of a dual AGN in Mrk~212. The optical emission line flux ratios obtained from the Himalayan Chandra Telescope (HCT) observations however, show that S1 and S2 both fall in the AGN + SF (star formation) region of the BPT diagram. Weak AGN lying in the SF or AGN + SF intermediate regions in the BPT diagram have indeed been reported in the literature; our sources clearly fall in the same category. We find an extended radio structure in our newly reduced 8.5~GHz VLA data, that is offset by $\sim1\arcsec$ from the optical nucleus S2. New deep FUV and NUV observations with the Ultraviolet Imaging Telescope (UVIT) aboard AstroSat reveal SF knots around S2 as well as kpc-scale tidal tails; the SF knots around S2 coincide with the extended radio structure detected at 8.5~GHz. The radio spectral indices are consistent with SF. Any possible association with the AGN in S2 is unclear at this stage.
%{\bf which could be associated with the AGN in S2 or the merging process.}}
\end{abstract}


\section{Introduction} \label{intro}
According to theories of galaxy formation, galaxies grow through mergers \citep{barnes1992}. Mergers are of two types: major mergers where the galaxies have a mass ratio of 1:3, and minor mergers where the mass ratio is less than 1:3. Major mergers have the ability to change the shape of the parent galaxies and form a galaxy with a new morphology \citep{BTbook2008} such as an elliptical galaxy whereas minor mergers mainly distort galaxies due to tidal forces and the associated star formation (SF) activity \citep{barnes1992}. Numerical simulations show that mergers induce strong gas inflows, which can give rise to nuclear starbursts \citep{mayer2007,karl2010}. We also know that almost all nearby massive galaxies host a super massive black hole at their centre \citep[SMBH;][]{kormendy1995,kormendy2013}. The gas inflow during mergers can trigger mass accretion onto the SMBHs thus turning them into active galactic nuclei \citep[AGN; e.g][]{sanders1988} and if both the SMBHs are ignited at the same time, AGN pairs are expected to form \citep{begelman1980}. These pairs are known as binary AGN if the separation is less than 100 pc and dual AGN for separations lying between 100 pc to 10 kpc \citep{burke2014}. However, several authors have obtained different definitions with different separation ranges \citep[e.g.][]{derosa2020}. Finally the two SMBHs coalesce, leading to the emission of gravitational wave radiation which could be detected using eLISA \citep{eLISA2013} and Pulsar Timing Arrays \citep[PTA;][]{manchester2013}.

A large sample of binary/dual AGN observed over a range of separations can help us to explore the following issues: (i) The end stage of the galaxy merging process when the SMBHs form a binary. Most binary AGN detections have been serendipitous, such as NCG~6240 \citep{komossa2003} and 3C~75 \citep{owen1985}. It is not clear where one must look for these candidates. (ii) How do the SMBH grow during mergers? (iii) How does the feedback associated with dual/binary AGN affect galaxy disks?

In recent years, double-peaked emission line AGN (DPAGN) have become popular dual/binary AGN candidates \citep[e.g.,][]{wang2009,smith2010,comerford2012,ge2012,fu2011}. However, DPAGN can also be due to jet kinematics \citep{kharb2015,kharb2019} or nuclear rotating disks \citep{mullersanchez2011,kharb2015}. 
 Many high resolution radio observations have been carried out to confirm dual/binary AGN in these DPAGN galaxies but this method has yielded very few confirmed dual or binary AGN \citep[e.g.,][]{fu2011a,mcgurk2015,mullersanchez2015,rubinur2019}. However, since dual/binary AGN arise from SMBHs in close interacting galaxy pairs that are simultaneously active \citep{liu2012,fu2018}, visually classified close galaxy mergers are also good dual/binary AGN candidates; the only drawback is that multi-wavelength observations are required to confirm the AGN nature of the nuclei. \citet{teng2012} observed 12 optically selected massive (10$^{11}$M$_{\odot}$) mergers with the Chandra X-ray observatory and \citet{fu2015, fu2015l} detected four dual AGN out of six merging galaxies using optical spectroscopy and radio imaging. Recent evidence shows that AGN are more obscured in galaxy mergers than in isolated galaxies \citep{liu2013,koceveski2015,ricci2015}. A good way around this problem is to select dual/binary AGN candidates using the mid-IR colors. \citet{satyapal2017} have confirmed four dual AGN in six sample galaxies. \citet{ellison2017} have discovered another dual AGN using high resolution Mapping Nearby Galaxies at Apache Point Observatory \citep[MaNGA;][]{bundy2015} survey using the same sample selection method.

One of the primary goals in detecting dual/binary AGN is to understand their effect on the host galaxies themselves. However, there are very few deep studies of double nuclei galaxies that also involve a detailed characterization of their AGN, the associated nuclear star formation and the feedback effects. Possibly because such studies require high-resolution, multi-wavelength observations. A good example of such a multi-wavelength study is by \citet{mazzarella2012} which investigates the confirmed dual AGN in Mrk~266. Recently, \citet{shangguan2016} have explored four dual AGN with high resolution optical and X-ray observations. However, they have not found any differences between the galaxy properties of dual AGN and a control sample of single AGN.
The ultraluminous infrared galaxy (ULIRG) NGC 6240 is one more such example. It is a gas rich merger remnant and has DAGN detected in X-ray to radio wavebands \citep[e.g.,][]{komossa2003}. Several studies have been carried out to trace the molecular and ionized gas outflows, and its connections between the two nuclei of NGC 6240 \citep[e.g.,][]{saito2018, muller2018}. Recently, the circumnuclear disk of this DAGN sytem was
mapped using the Atacama Large Millimeter/submillimeter
Array (ALMA)\footnote{https://www.eso.org/public/teles-instr/alma/} which provided the first measurement of significant molecular gas masses
contaminating dynamical black hole mass measurements \citep{medling2019}.

Star formation rates (SFRs) are important tracers that can be used to  understand the evolution history of a galaxy as well as the connection between SMBHs and galaxy growth \citep{kormendy2013, alexander2012}. It can be measured at different wavelengths. These different wavelengths may, however, lead to discrepancies in the estimated SFR owing to the sensitivity to dust extinction and irregular star formation history \citep{calzetti2013}. H$\alpha$ traces massive ($\ge$17 M$_{\sun}$) O and B type stars and recent star formation of age a few million years. The ultra-violet (UV) emission measures older star-formation that have age of tens of million years. The far-UV (FUV) traces the massive stars (e.g., O, B types) while the near-UV (NUV) traces less massive stars (e.g., A, B types) which are more common. The dust present in the stellar regions absorbs the radiation and re-emits in infrared (IR) light. Massive stars explode at the end of their life and turns into supernova which dominates at the non-thermal radio emission at low frequencies ($\le$ 5 GHz) \citep[e.g.,][]{Kennicutt.etal.1998, condon1992}.

\begin{figure}
%\hspace{1cm}
\includegraphics[trim= 0cm 0 0cm 0, width=1.1\columnwidth]{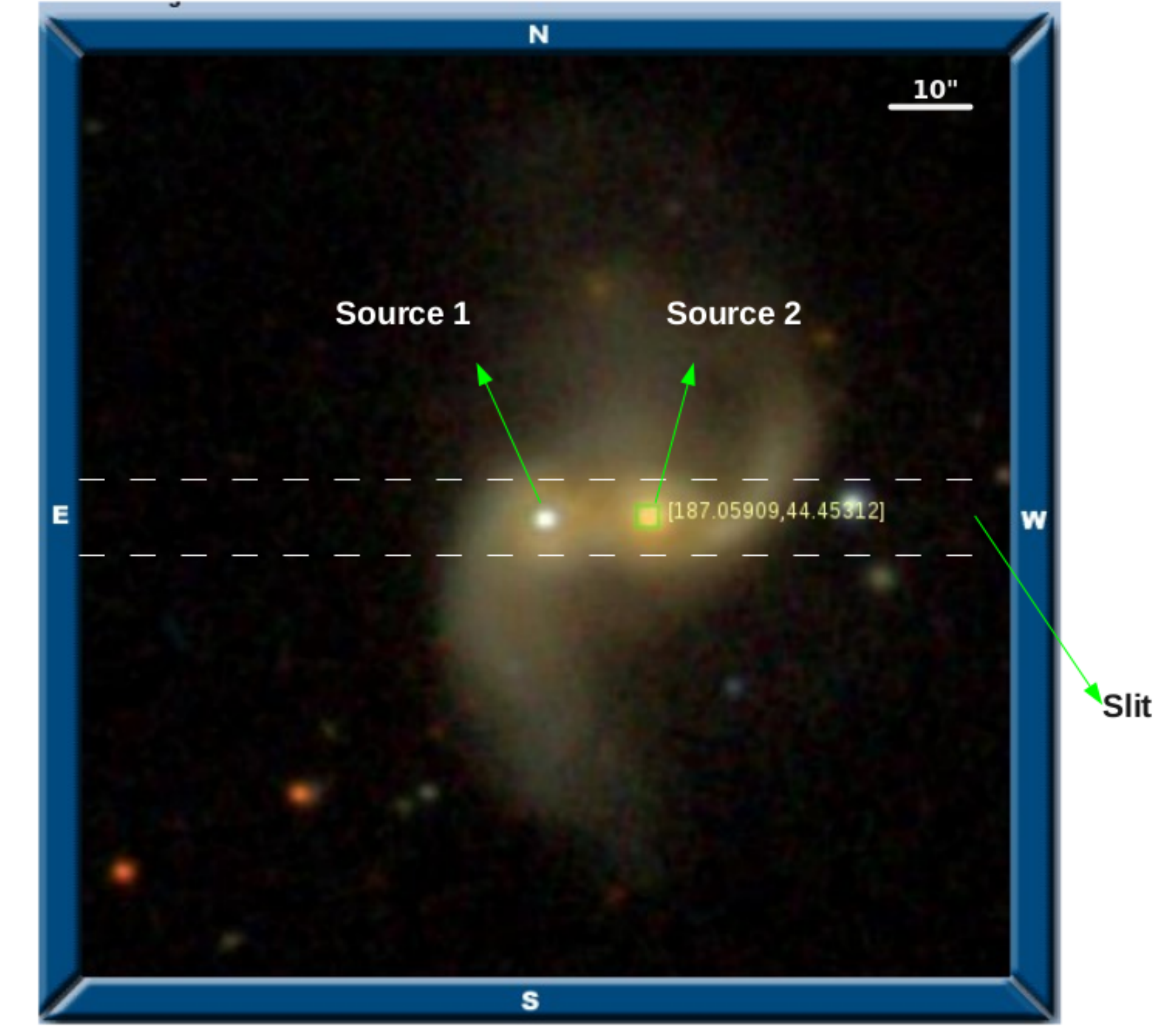}
\caption{The SDSS color composite image of Mrk~212. The slit position of the HCT observations is indicated by the dashed line. Mrk~212 is referred to as Source 1 (S1) and the companion as Source 2 (S2).}
\label{sdss}
\end{figure}

In this paper, we present results from our deep radio, optical and UV study of the dual nuclei in the galaxy merger  Mrk~212. It is a double-pinwheel galaxy (UGC 07593 NED02) where the companion galaxy is UGC~07593 NED01. In this paper, we refer to Mrk 212 as Source 1 (S1) and its companion as Source 2 (S2) (see Figure \ref{sdss}). S1 has a redshift ($z$) of 0.022893 $\pm$ 0.000133 \citep{rc1991}
%(luminosity distance D$_L$ = 98.8 Mpc, scale = 0.459 kpc/arcsec)\footnote{https://ned.ipac.caltech.edu/}. 
S2 has a $z$ of 0.022352 $\pm$ 0.000063. 
%(luminosity distance D$_L$ = 96.5 Mpc, scale= 0.449 kpc/arcsec).
The luminosity distance of the system is adopted as D$_L$ = 98.8 Mpc, scale = 0.459 kpc/arcsec.
The Sloan Digital Sky Survey \citep[SDSS;][]{blanton2017}\footnote{http://www.sdss.org/} image (Figure \ref{sdss}) shows that the projected nuclear separation is 11.8 arcsec or $\sim$5.6 kpc. The two radio sources are resolved in the Faint Images of the Radio Sky at Twenty cm \citep[FIRST;][]{becker1994}\footnote{http://www.cv.nrao.edu/first/} image 
%(Figure \ref{first}) 
but not in the NRAO VLA Sky Survey \citep[NVSS;][]{condon1998}\footnote{http://www.cv.nrao.edu/nvss/} image. Both sources are spiral galaxies of type Sb and Sbc. The SDSS image shows that the inner arms of each galaxy are nearly merged, while the outer arms appear to be unaffected (Figure \ref{sdss}). \citet{mezcua2014} have carried out a photometric analysis of this system and obtained r-band magnitudes of 21.77 and 22.86 respectively, for S1 and S2 respectively. There is an SDSS optical spectrum of S2 and it is identified as a star-forming galaxy by SDSS DR12\footnote{https://www.sdss.org/dr12/}. 

One of the aims of this study was to investigate the nature of the nuclear activity in S1 and S2 (AGN or SF).
%, whether they are AGN or star forming in nature. Hence, see if Mrk212 is a DAGN. But apart from that, we also wanted to study the effect of the dual nuclei on the galaxy itself. Basically 
We also wanted to study the effect of the dual nuclei on the merging galaxy itself and look for signatures of AGN/SF feedback.
%the AGN/SF feedback from the nuclei at kpc scales and its signatures at different wavelengths.} 
In the following sections, we present our observations of Mrk~212, the results and discuss the implications of our work for the detection of DAGN in this double nuclei galaxy. Throughout this paper we assume a value of $\Omega_{m}=~0.27$, and $H_{0}=~73.0$~km~s$^{-1}$~Mpc$^{-1}$.
The spectral index, $\alpha$, is defined such that the flux density at frequency $\nu$ is S$_\nu\propto~\nu^\alpha$.

\section{Observations and Data Analysis}
\subsection{Radio data} \label{radio_data}
We first analysed VLA archival data from the NRAO data archive: (i) Mrk 212 was observed at 4.86 GHz using the B-array configuration on 14th July, 1986 (Project ID: AM183) for $\sim$11 min along with the flux density calibrator 3C286 for 3 min. J1323+321 was observed for 2 min as a phase calibrator. It is a single channel observation with a total bandwidth of 50 MHz. This data is in B1950 epoch. (ii) Mrk~212 was observed at 8.5 GHz using the VLA BC-array on 16th Feb, 2004 (Project ID: AB1122). The science target was observed for 4 min along with flux density calibrator, J1331+305 for 6 min and phase calibrator J1219+484 for 5 min. The total bandwidth was 50 MHz.

After obtaining images at 4.86 and 8.5 GHz, we observed Mrk 212 on 17th Sep, 2017 (Project ID: 17-123) with the VLA in the Ku-band in the B-array. The observations were carried out with 17 spectral windows (spw), each having 64 channels i.e a total bandwidth of 1.28 GHz. Mrk 212 was observed for 8.30 min. The flux density calibrator 3C286 and the phase calibrator J1219+4829 were observed for 5 min and 3.30 min respectively. We have obtained uGMRT observation of Mrk 212 on 4th Nov, 2018 with band 5 at 1.4 GHz. The uGMRT band 5 has a total of 2047 channels with a bandwidth of 400 MHz. The science target was observed for $\sim$4 hrs and the flux density calibrator, 3C286 and phase calibrator, J1219+4829 were observed for 20 min and 65 min respectively.

We have used the Common Astronomy Software Applications \citep[CASA;][]{mcmullin2007} and Astronomical Image Processing System \citep[AIPS;][]{moorsel1996} packages for the VLA and GMRT data reduction as well as for further image analysis. First, we identified bad data (radio frequency interference, zero amplitude data etc) in the calibrators using the {\sc plotms} interface in CASA, and flagged them. The model for the flux calibrators was defined using the task {\sc setjy}. The initial phase calibration was carried out using the task {\sc gaincal}. Subsequently, the delay and bandpass corrections were obtained. These solutions were applied and followed by the gain calibrations of the flux and phase calibrators. 

Finally, the calibrations were applied to the science target. This whole process was repeated until a satisfactory calibration was obtained. The {\sc clean} task was used to make the image of the science target. We have also used images from the NRAO archival image database, where available, for comparison and further analysis. The {\sc jmfit} task was used to obtain the flux density and the errors. The spectral index images were made using the {\sc comb} task in AIPS. Images at both frequencies were convolved with a similar beam-size at an intermediate resolution and made coincident using the task {\sc ogeom}.

\begin{center}
\begin{table*}
\caption{The radio observation details}
\begin{tabular}{lcccccc}

\hline
Band    & Central frequency (GHz) & Array & Resolution & Exposure time & Date of observation & Observation ID\\
\hline

% L-band & 1.4 & A & 5.00$^{\prime\prime}$  \\
C-band &  4.86 & B &1.46$^{\prime\prime}$& 12 min & 14th July, 1986 &VLA/AM183 \\
X-band &  8.46 & BC &1.03$^{\prime\prime}$ & 4 min & 16th Feb, 2004 & VLA/AM1122 \\
Ku-band & 13.5 & B & 0.40$^{\prime\prime}$&12 min & 8th Sept, 2017 & VLA/17B-123  \\
Ku-band & 16.5 & B & 0.40$^{\prime\prime}$&13 min & 8th Sept, 2017 & VLA/17B-123 \\
L-band & 1.46 & - & 2.0$^{\prime\prime}$& $\sim$ 4 hrs & 5th Nov, 2018 & uGMRT/35-076 \\ \hline
\label{tab_rad}                                                               
\end{tabular}
\end{table*}
\end{center}

\subsection{Optical spectroscopy data}\label{opt_data}
The optical spectroscopic observations were done using the HFOSC instrument mounted on the 2-m Himalayan Chandra telescope (HCT) located at Hanle, India. The slit has a dimension of 11$^{\prime}\times 1.92^{\prime\prime}$ and was used in combination with grism 7 for the observations. A grism is a dispersive element mounted on the telescope and is a combination of a transmission grating and a prism. The grism that we used has a dispersion of 1.46\AA\ and covers a wavelength range of 3700$-$7200\AA. The observational details are presented in Table~\ref{tab_optobs}. The slit was positioned so that it covered both the nuclei. The wavelength calibration was carried out using the ferrous argon (FeAr) arc lamp and the flux was calibrated using the standard stars from \citet{oke1990}. We have used HZ~44 as the standard star.
\par The data reduction was done using the software package {\sc IRAF} (Image Reduction \& Analysis Facility).
The pre-processing steps include trimming and bias subtraction. The tasks {\sc imcopy, zerocobime} and {\sc imarith} were used. The one dimensional spectrum was extracted using the task {\sc apall}. This task includes defining the aperture, the spectrum extraction and tracing the spectrum along the dispersion axis. The task can handle data which is in multi-spectral format and over different bands. Using the resultant multi-spectral data, the band corresponding to the spectrum of the galaxy is separated out. It is referred to as the galactic spectrum. The error spectrum corresponding to each source is also extracted. For the separation of different bands the task {\sc scopy} was used. The spectrum from each individual nucleus was extracted by defining suitable apertures. The calibration lamp spectra were extracted with reference to the respective target sources. The atomic emission lines in the calibration lamp spectra were identified and the pixel to wavelength mapping was done using the task {\sc identify} and {\sc dispcor}, respectively. The wavelength calibrated spectra were also flux calibrated using the tasks {\sc standard} and {\sc calibrate} with the standard star spectrum taken on the same night. The resolution was determined with the skyline at 5577 \AA.  The full width at half maximum (FWHM) obtained at 5577 \AA\ is $\sim$ 8.5 \AA\ for grism 7. The flux calibrated spectra were redshift corrected using the task {\sc dopcor} and corrected for reddening due to the Galaxy. 

%\begin{center}
\begin{table}
\caption{HCT optical observation details}
%%%\scriptsize
\begin{tabular}{lcc}

\hline
Source     &    Exposure time(s)& Date of observation\\
\hline

Mrk 212  &   2700    & 2019-05-29 \\
HZ 44    &   900 	 & 2019-05-29 \\
\hline                            
\label{tab_optobs}                                                               
\end{tabular}
\end{table}
%\end{center}

\subsection{X-ray archival data}\label{x_data}
The X-ray data on Mrk 212 are available in the Chandra data archive\footnote{https://cxc.harvard.edu}. The observations were carried out with the Advanced CCD imaging spectrometer (ACIS-I) on 10th April, 2000 (PI: Takahasi) and were aimed at observing the X-ray source AX J1227.6+4421. Mrk~212 is situated 7$^{\prime}$ away from the target source. The total integration time of the observation is 5.65 ks. The expected spatial resolution is $\sim1\arcsec$. The archival primary event files were reprocessed following the standard procedure using \textit{chandra\_repro} task within \textsc{CIAO} v4.12. An annular region surrounding the double nuclei was selected for background subtraction. The observation was energy filtered with range 0.3$-$7 keV. Standard Chandra data-reduction threads were followed for the analysis and the details given in Section \ref{x-ray}.

\subsection{UV data}\label{uv_data}
We have observed Mrk 212 with the Ultra-violet imaging telescope (UVIT). The UVIT is one of the five payloads onboard the AstroSat satellite \citep{kumar2012} launched by ISRO\footnote{https://www.isro.gov.in/} on 28th Sept, 2015. It consists of two co-aligned UV telescopes with the field of view of 28$\arcmin$ and a spatial resolution of $\sim$ 1.5$\arcsec$, which is three times better than the GALEX resolution of $\sim5\arcsec$. The telescopes have an aperture size of 375 mm. Observations are done in the three bands - FUV (1300$-$1800~\AA),  NUV (2000$-$3000~\AA) and visible (VIS) bands simultaneously. The FUV, NUV and VIS channels consist of multiple filters of different bandwidths which allows for narrow, medium and broad band imaging. The VIS channel is used for drift correction induced during the observations. Mrk 212 was observed with the UVIT in two cycles. Here, we present the longer duration 15 ks UVIT observation of Mrk 212 (A04-091). The observations were done using the FUV CaF2 (1481- 1981 \AA) and NUV-Silica (2418- 3203 \AA) broad band filters. We have reduced the data using the JUDE (Jayant's UVIT data explorer) pipeline \citep{murthy2016,murthy2017}. The UVIT counts per second (CPS) were converted to flux using calibration factors from \citet{rahna2017}. The UVIT pipeline images were also used in the analysis.

\begin{figure*}
\includegraphics[width=18cm,trim=20 280 20 280]{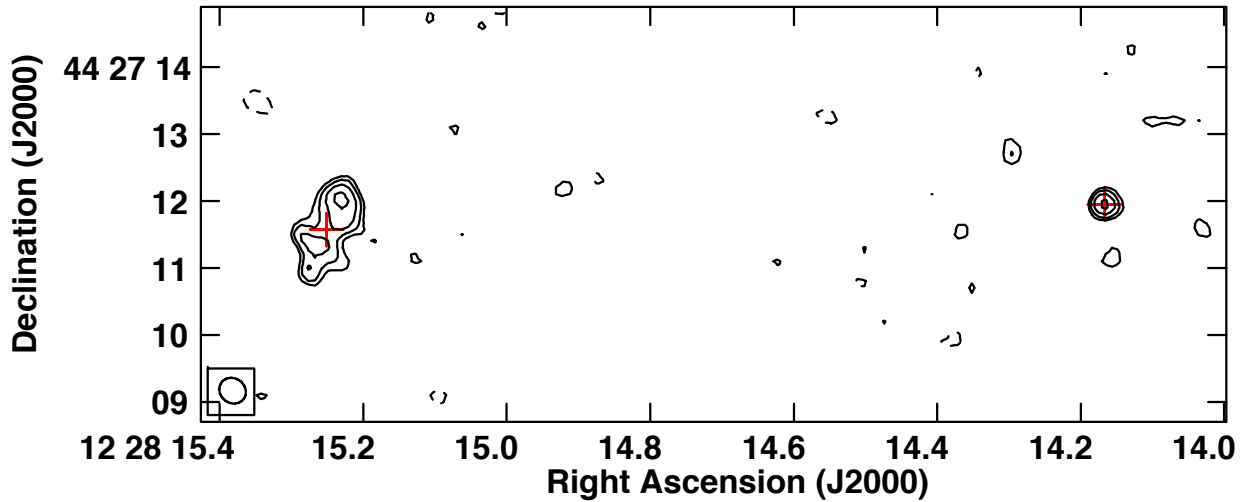}
\caption{15~GHz VLA contour image of Mrk~212. The optical centres of S1 and S2 are obtained from the Pan-STARSS1 r-band image and marked with red crosses. The contours levels are: -32, 32, 45, 64, 90\% of the peak surface brightness of 5.6$\times10^{-5}$~Jy/beam.}
\label{ku_pan}
\end{figure*}

\begin{figure*}
\includegraphics[width=15cm,trim=0 0 0 10]{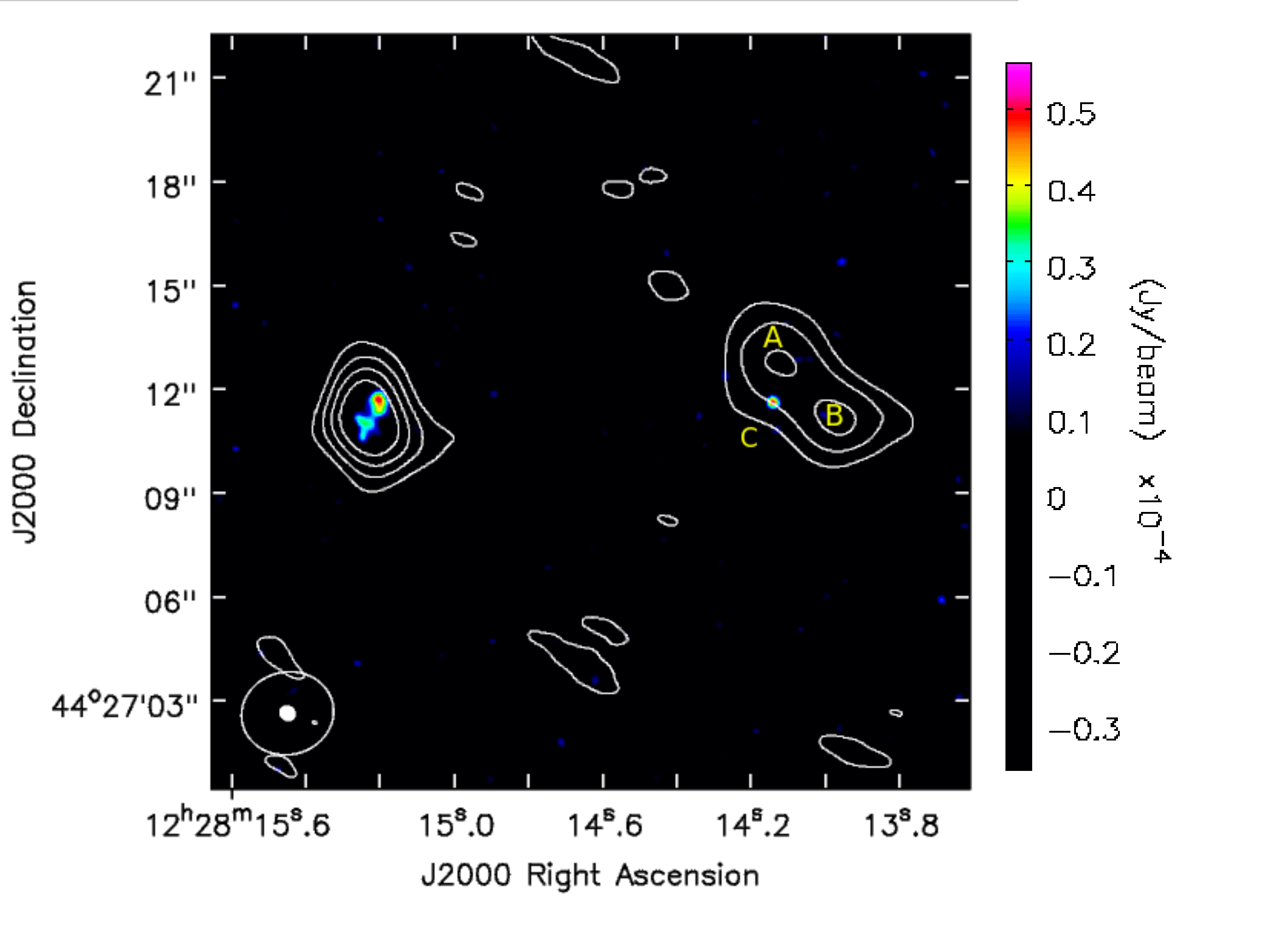}
\caption{8.5~GHz VLA contour image showing radio emission associated with S1 and S2. The 15~GHz EVLA image is superimposed in color. S2 shows an offset extended structure, marked with features A and B. C marks the position of the point source associated with S2. The contour levels are 20, 40, 60 and 80\% of the peak flux density value 0.54 mJy/beam.}
\label{x_band}
\end{figure*}

\begin{figure*}
\includegraphics[width=12cm,trim=10 50 10 10]{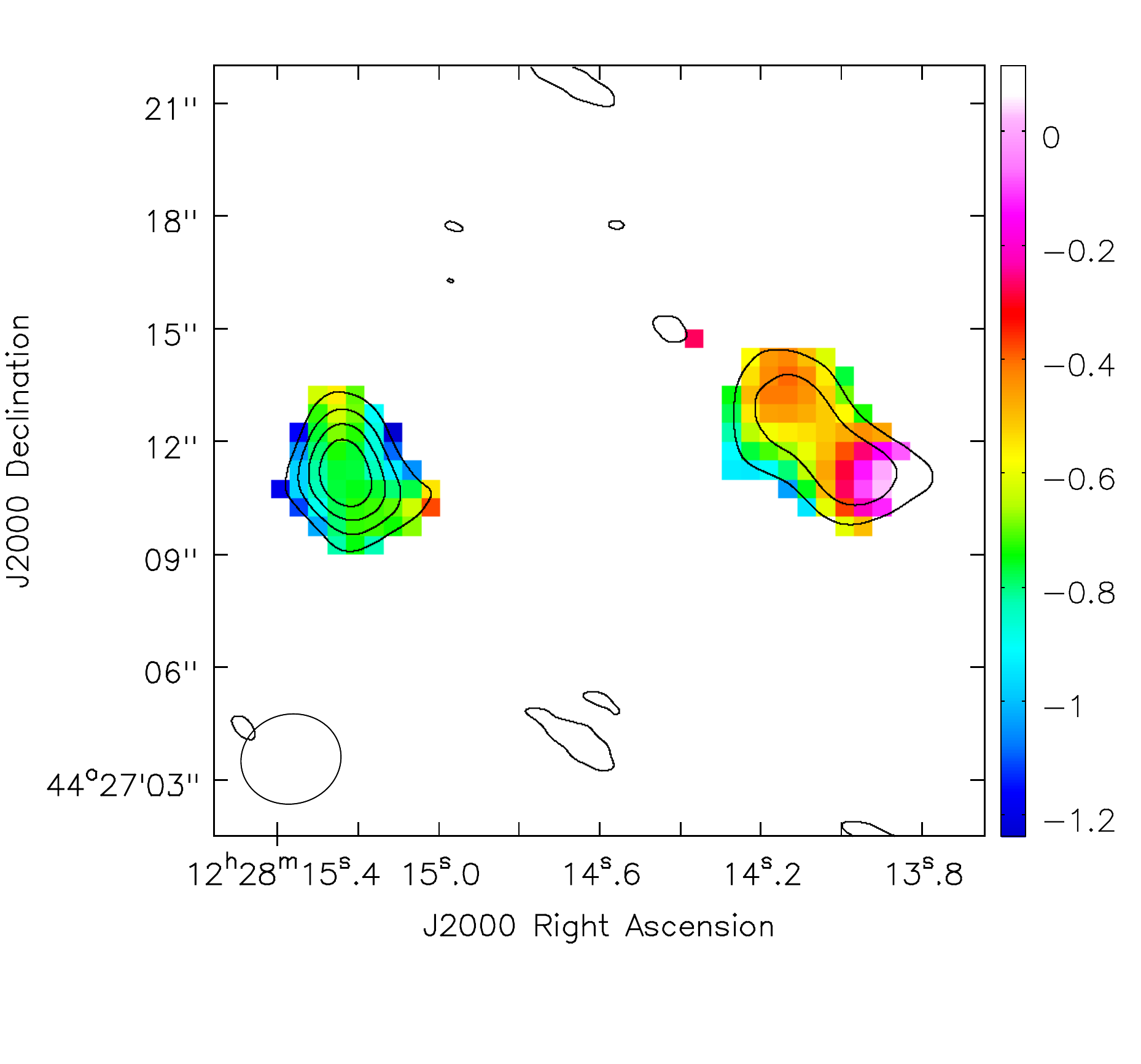}
\caption{The spectral index image from the VLA 8.5 GHz and uGMRT 1.4 GHz images. The 8.5 GHz contours are overlaid. The contour levels are 20, 40, 60 and 80\% of the peak flux density value.}
\label{sp_gmrt}
\end{figure*}

\section{Results}
\subsection{Radio Properties}\label{radio}
We have analysed archival 4.86 GHz, 8.5 GHz and new 15 GHz VLA data, as well as 1.4 GHz uGMRT data of Mrk~212. As mentioned earlier, the VLA FIRST image shows two radio sources coincident with the optical nuclei, S1 and S2 (Figure~\ref{first}). Our new observations at 15 GHz resolve the radio source coincident with S1 into a double-lobed source and S2 into a compact core (Figure~\ref{ku_pan}). Below we discuss the results at individual frequencies. 

(i) The 4.86 GHz image shows a single radio source at the position of S1. No emission is detected coincident with S2. This is due to lack of adequate sensitivity in the data. The natural weighted integrated flux density is 0.41 mJy. The task {\sc imregrid} in CASA was used to transform the image from B1950 to the J2000 coordinate system. 

(ii) The 8.5 GHz image (Figure~\ref{x_band}) shows two radio structures at a separation of $\sim$ 6 kpc. The S1 shows a double radio source with an extended structure in the North and South direction. An extended radio structure of size $\sim$2 kpc is detected 1$\arcsec$ from S2. It has two radio peaks (components A, B in Figure \ref{x_band}) in the North East - South West direction. 

(iii) The 15 GHz image (Figure \ref{ku_pan} and \ref{x_band}) shows an extended radio structure associated with S1 and a compact radio core (C) associated with S2. The extended structure in S1 resembles a double-lobed radio source. The radio core C is offset from the optical centre of S2 (in the Pan-STARRS1 r-band image) by 0.36$\arcsec$ in RA and 0.31$\arcsec$ in DEC. Such offsets are typical between optical and radio images and are likely due to residual phase errors in the radio images. Once we correct for this offset using the {\sc ogeom} task in AIPS, the optical centre of S1 falls smack in the middle of the two lobes in the extended structure (see Figure \ref{ku_pan}). The total size of this structure is $\sim$750 parsec, making it similar to a compact symmetric object (CSO) \citep{perucho2002}. The deconvolved size of the unresolved component C using the {\sc JMFIT} task in AIPS is $\sim$57~parsec. Since this component is not picked up clearly in the 8.5 GHz image, we used the peak flux density of C to calculate the expected flux density at 8.5 GHz using an $\alpha=-0.7$. This gives an expected value of the flux density of 69.8 $\mu$Jy which is comparable to the r.m.s. noise in the 8.5 GHz image, explaining the lack of a clear detection. On the other hand, component C could have an inverted spectrum, consistent with it being the synchrotron self-absorbed base of an AGN jet, which could explain why it becomes prominent at the highest observed radio frequency.

(iv) {Figure \ref{gmrt}} shows the uGMRT image of Mrk~212 at 1.4~GHz. The uniformly weighted image shows two radio sources. Both the sources have some extended emission of the order of 3$\arcsec$ to 4$\arcsec$ (Figure \ref{gmrt}).

% \onecolumn
\begin{table*}
\centering
\caption{{\bf Radio properties of Mrk 212:} column 1: Observation ID; column 2: frequency of the observation in GHz; column: 3: beamsize with major axis ($\theta_{1}$), minor axis ($\theta_{2}$) and position angle (PA)); column 4: rms noise in the radio images in $\mu$Jy; column 5: integrated flux density of the S1 in mJy; column 6: integrated flux density of the S2 in mJy; column 7: separation between the sources in arcsec; column 8: separation between the sources in kpc. {\bf Note 1:} $^\star$ the second source is not detected in the 4.86 GHz image. {\bf Note 2:} The separations are calculated with respect to optical centers. For 8.5 GHz image, the middle of the extended structure of S2 is considered.}
%\vspace{0.8cm}
%\normalsize
\label{tab_radresult}
\resizebox{1.0\textwidth}{!}{
\begin{tabular}{|c|c|c|c|c|c|c|c|c|c|}
%\tableline\tableline\begin{tabular}[c]{@{}l@{}}Observing \\Frequency \end{tabular}
\hline
Observation ID &\begin{tabular}[c]{@{}l@{}}Frequency \\ (GHz) \end{tabular}
&\begin{tabular}[c]{@{}l@{}}Beam \\ ($\theta_{1} \times \theta_{2}$, PA) \end{tabular} 
& \begin{tabular}[c]{@{}l@{}} Image  \\ Noise \\ ($\mu$Jy) \end{tabular}
& \begin{tabular}[c]{@{}l@{}} S1 \\ Int Flux \\ density (mJy)\end{tabular}
&\begin{tabular}[c]{@{}l@{}}S2\\ Int Flux\\ density (mJy)\end{tabular}
&\begin{tabular}[c]{@{}l@{}} Separation \\ (arcsec) \end{tabular}
&\begin{tabular}[c]{@{}l@{}} Separation \\ (kpc) \end{tabular} \\\hline
FIRST         & 1.40  &   5.40$^{\prime\prime}\times$5.40$^{\prime\prime}$, 0.0$^\circ$    &165 & 2.02$\pm0.42$ & 2.35$\pm0.25$ & 12.79$^{\prime\prime}$ &5.84 \\ 
VLA/AM183     & 4.86  &   2.50$^{\prime\prime}\times$1.00$^{\prime\prime}$, 67.27$^\circ$  &55   & 0.41$\pm0.06$ & -$^\star$ &- & -  \\ 
VLA/AB1122    & 8.46  &   2.67$^{\prime\prime}\times$2.38$^{\prime\prime}$, -77.73$^\circ$  &65   & 0.74$\pm0.06$ & 0.56$\pm0.08$ & 13.06$^{\prime\prime}$ & 5.96 \\ 
VLA/17B-123   & 14.99 &   0.41$^{\prime\prime}\times$0.36$^{\prime\prime}$, 52.65$^\circ$  &6    &0.30$\pm0.05$  &0.05$\pm0.01$ & 11.33$^{\prime\prime}$ & 5.17\\ 
GMRT/35-076   &1.40   &   2.67$^{\prime\prime}\times$2.38$^{\prime\prime}$, -77.73$^\circ$  &15   &3.43$\pm0.19$  &2.28$\pm0.27$  & 12.44$^{\prime\prime}$ &5.68 \\ \hline
%\tableline 

\end{tabular}
}
\label{table3}
\end{table*}

\begin{figure*}
\includegraphics[width=12cm]{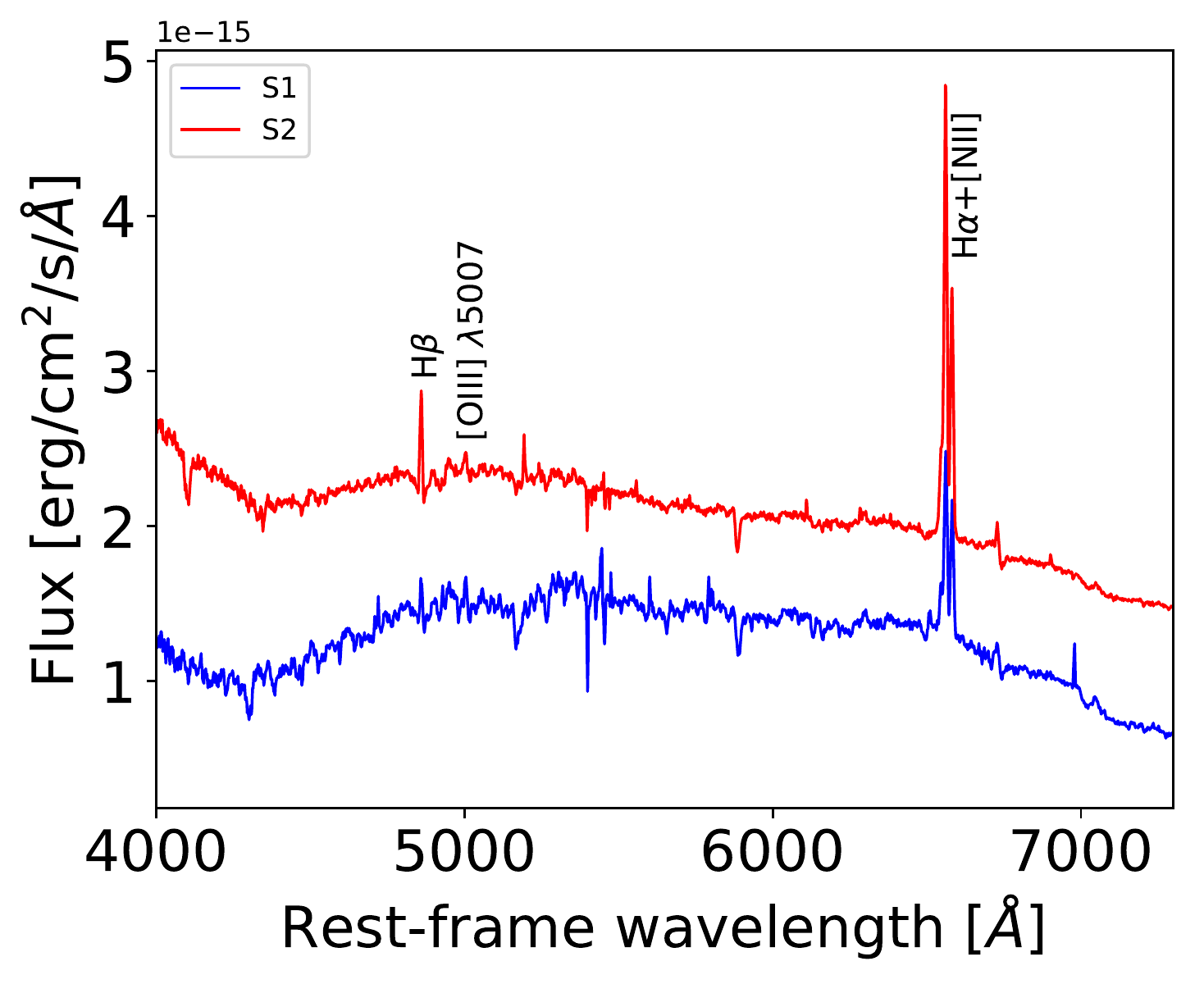}
\caption{The optical HCT spectra of S1 (blue) and S2 (red), where some of emission lines used in this work are identified and marked.}
\label{bestfitspec}
\end{figure*}

%\subsection{Radio spectral index}\label{sp_map}
We have created a spectral index image using the 1.4~GHz uGMRT and 8.5~GHz VLA images (Figure \ref{sp_gmrt}). The 8.5~GHz image is convolved with the restoring beam of the uGMRT image using the {\sc clean} task. These are discussed ahead in Section~4. We have not used 4.86 GHz image to create the spectral index map as it does not show both the nuclei.

\subsection{Optical spectra}\label{optical}
The optical spectra of the individual sources (i.e., S1 and S2) were obtained using the HCT, which are shown in Fig \ref{bestfitspec}. The characteristic AGN emission lines are H$\alpha$, [NII]$\lambda$6583, [OIII]$\lambda$5007 and H$\beta$. The fluxes for these emission lines were measured with the {\sc splot} task available in {\sc IRAF}. For regions with multiple emission lines such as the H$\alpha$+ [NII] region, a simultaneous multi component model fit is required. Hence, the H$\alpha$+ [NII] region was fitted with three gaussian profiles (two for the [NII] doublets and one for the narrow line component of H$\alpha$) with the de-blend procedure in {\sc splot} task. The {\sc splot} task also subtracts the underlying local continuum flux in the emission line flux measurements. The errors in the emission line flux measurements were estimated using $rms$ $\times$ $\sqrt{2 \times N}$, where $N$ is the number of pixels covered in the Gaussian profile of the emission line. The $rms$ was estimated from the line-free region (continuum) on both sides of the emission line. 

In some cases, in addition to a narrow component, the H$\alpha$ line may have a broad component as well, which arises from the broad line region (BLR) clouds.  However, for both of the sources in Mrk~212 broad line components of H$\alpha$ were not prominently seen. Similarly for the H$\beta$ line there is no broad line component, and so they are fitted with one narrow line component.

\begin{figure*}
\includegraphics[width=12cm]{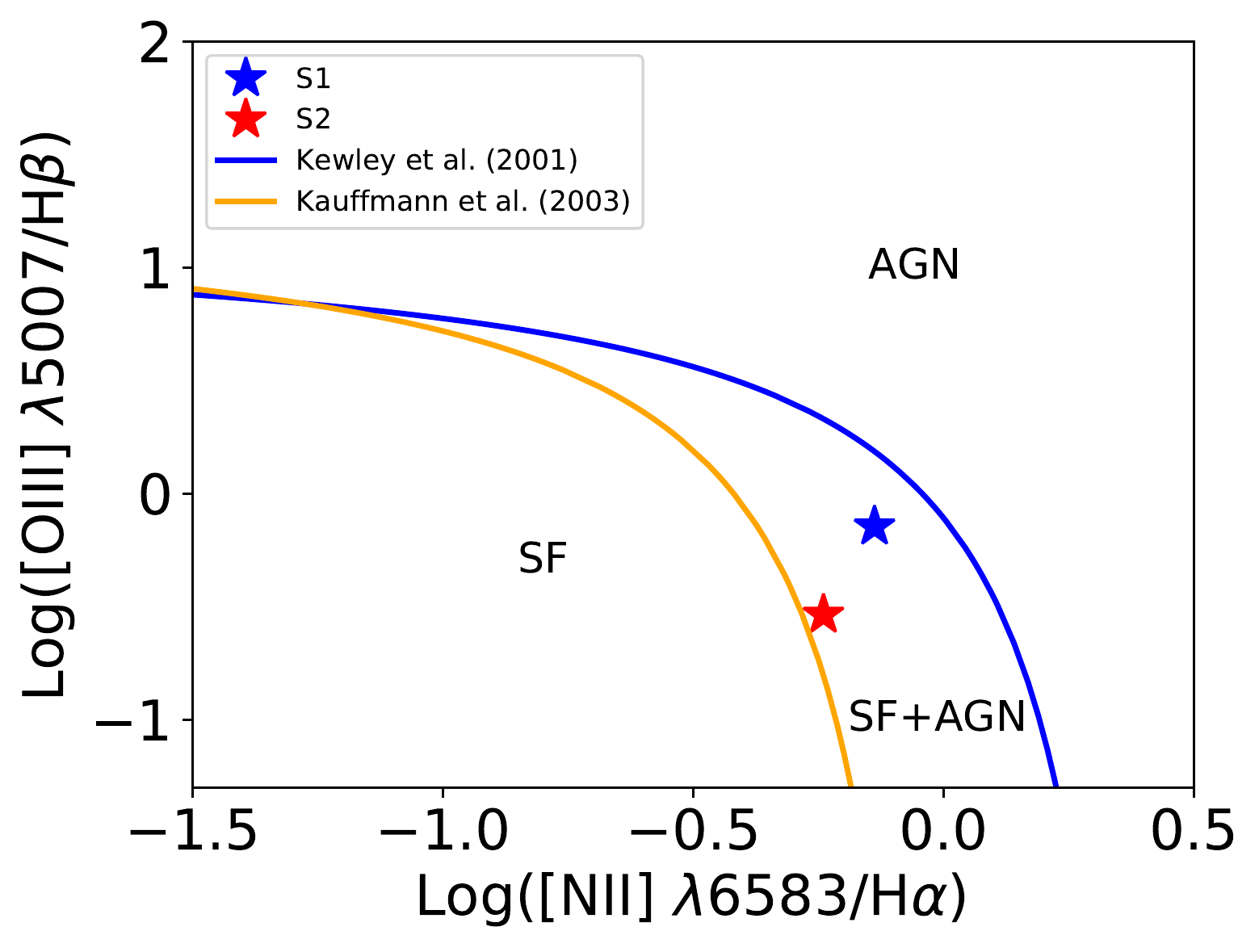}
\caption{The BPT diagram with source positions. The blue star is S1 which falls in composite (SF+AGN) region. Red star represents S2 which also falls in composite region.}
\label{figone}
\end{figure*}
 
The obtained emission line fluxes are given in Table \ref{tab_optical}. We have calculated the ratio of [NII]$\lambda$6583 to H$\alpha$ and [OIII]$\lambda$5007 to H$\beta$ for both sources. The points were plotted in the Baldwin, Philips and Terlevich \citep[BPT;][]{BPT1981} diagram (see Fig \ref{figone}) which is an empirical diagram with the boundaries from \citep{kewley2001} and \cite{kauffmann2003}. This is one of the best ways in the optical waveband to distinguish between AGN (Seyfert and quasar), LINERs and HII regions. In Fig \ref{figone}, our BPT diagnostic suggests that both the sources have a composite nature i.e., a mixture of AGN and SF activities responsible for emission line excitation.     

 \begin{table}
 \centering
 \caption{{\bf Fluxes of the best fitted optical emission line in the units of $10^{-15}$ ~erg~cm$^{-2}$~s$^{-1}$:}}
  \hspace{-1cm}
\vspace{-0.5cm}
\label{tab_optical}
\resizebox{0.5\textwidth}{!}{
\begin{tabular}{ccccccc}
\hline Galaxy        &  \begin{tabular}[c]{@{}c@{}}H$_\beta$\\ 4861 \AA \end{tabular} & \begin{tabular}[c]{@{}c@{}}H$_\alpha$\\ 6563 \AA \end{tabular} & \begin{tabular}[c]{@{}c@{}} [O III]\\ 5007 \AA \end{tabular} & \begin{tabular}[c]{@{}c@{}}[N II]\\ 6583 \AA \end{tabular} \\ \hline

S1 &  2.65 $\pm$ 0.27 & 14.3 $\pm$ 0.2 & 1.90 $\pm$ 0.24 & 1.04 $\pm$ 0.02\\
S2 & 5.59 $\pm$ 0.12 & 32.7 $\pm$ 0.1 & 1.63 $\pm$ 0.14 & 18.8 $\pm$ 0.1\\ \hline
 
\vspace{0.2cm}
\end{tabular}}
\end{table}
\vspace{2cm}

\subsection{X-ray emission}\label{x-ray}
The spectral fitting is carried out using the \textsc{sherpa} fitting tool. The source is fitted with a basic photon power-law and photo-electric absorption whose equivalent hydrogen column is determined from the Heasoft nH tool. The estimated integrated flux in the energy range 0.3$-$7 keV is 3.4 $\times$ 10$^{-14}$~erg~s$^{-1}$~cm$^{-2}$ and the best fit photon index is $\Gamma = 2.48^{+1.21}_{-0.99}$. We also measured count rates and determined the hardness for each nucleus (tabulated in Table \ref{tab:count_rates}). The regions used for measuring count rates are marked in Figure \ref{dss_regions}. These regions enclose S1 and S2. The soft band rate for S1 is twice that of S2 while the hard band rate for S2 is a factor of three higher than S1. This suggests that S1 has a softer spectrum while S2 has a harder spectrum. The hardness ratio is defined as HR = (H-S)/(H+S) where S and H are the soft (0.5$-$2 keV) and hard (2$-$7 keV) X-ray band net counts respectively. The HR for S2 is higher compared to S1 indicative of heavy obscuration, which could be due to an increase in the local absorption column density in merging systems \citep{kocevski2015,ricci2017}. Due to the low number of counts, an additional intrinsic absorption component to the power-law fit resulted in poor statistics and the parameters could not be constrained. Hence, the X-ray data is not sufficient to distinguish between low luminosity AGN from a nuclear X-ray binary \citep{she2017}. Deeper X-ray observations are needed for confirmation.

\begin{table}
		\caption{Count rates for the soft band (0.5$-$2 keV) and hard band (2$-$7 keV) for regions defined in Figure ~\ref{dss_regions}.}
	\begin{tabular}{cccc}
		\hline
		Region	&	Soft rate (S)	&	Hard rate (H)	&	Hardness Ratio \\
				&	(10$^{-3}$~ct~s$^{-1}$)	&		(10$^{-3}$~ct~s$^{-1}$)	&	(H-S)/(H+S))\\
		\hline
		S1	&	1.61$\pm$0.54	&	0.36$\pm$0.25	&	-0.63$\pm$-0.23\\
		S2	&	0.72$\pm$0.36	&	1.26$\pm$0.45	&	0.27$\pm$0.28\\
		\hline			
	\end{tabular}
		\label{tab:count_rates}
	\end{table}
	
\subsection{UV emission}\label{uv}
\begin{figure*}
\centerline{\includegraphics[width=12cm, trim=0 10 0 0]{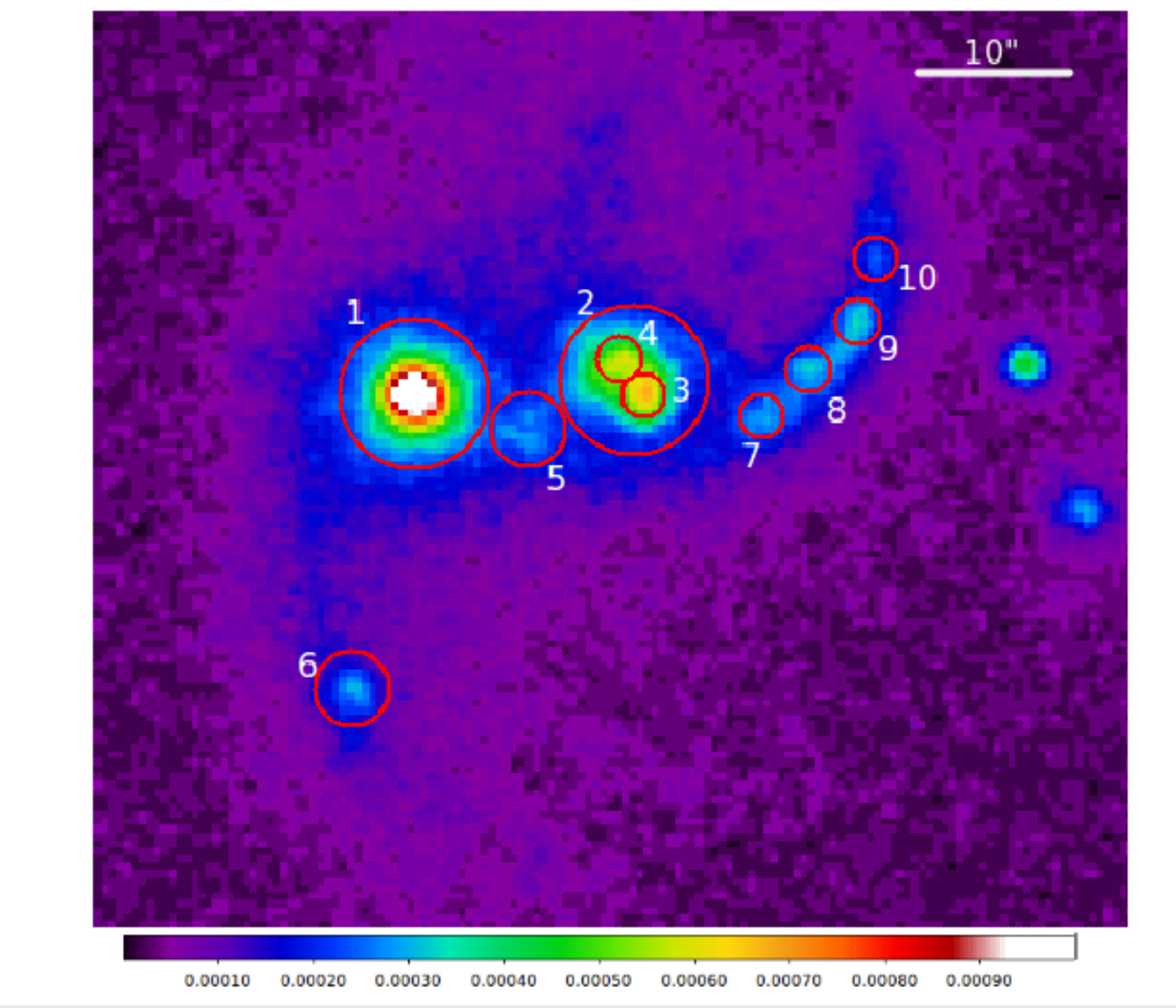}}
\caption{The NUV image of Mrk 212 from UVIT observation. This deep 15 ks image has resolved star-forming knots. S2 has two such bright UV knots (knots 3 and 4) which coincide with the X-band two sided structure (A, B). The circles are regions selected to calculate the SFRs (see Table~6).}
\label{nuv}
\end{figure*}

The GALEX image of Mrk~212 shows UV emission from the two nuclei as well as from the tidal arms. The high-resolution AstroSat-UVIT image with an exposure time of 2.5 ks could only resolve the UV emission lying between the two galaxies. However, in the longer exposure (15 ks) image (Figure \ref{nuv}), we have detected emission from both galaxies, the tidal arms as well as from the star-forming UV knots in S2. The two UV knots in S2 (region 3 and 4 in Figure \ref{nuv}) are resolved and they coincide with the regions A and B in the 8.5 GHz image. We have visually selected the UV bright regions and calculated the SFR using \citet{calzetti2007} while following \citet{rahna2018}. The following relation is used \citep{calzetti2007}: 
\begin{equation}
\mathrm{SFR}_\mathrm{UV} (\mathrm{M}_{\sun}~\mathrm{yr}^{-1})= \mathrm{C}\times \mathrm{L}_\mathrm{UV}\times\lambda_\mathrm{UV} \nonumber
\end{equation}
Here, C = 3.0$\times 10^{-47}$ for the FUV and C = 4.2$\times 10^{-47}$ for the NUV emission, respectively, L$_{UV}$ is the UV luminosity and $\lambda_{UV}$ is the UV wavelength. The calculated SFR ranges from 1$\times 10^{-3}$ to 96$\times 10^{-3}$ M$_{\odot}$~yr$^{-1}$ in the NUV and 0.7$\times 10^{-3}$ to 34$\times 10^{-3}$ M$_{\odot}$~yr$^{-1}$ in the FUV. The SFR$_\mathrm{NUV}$ is higher than SFR$_\mathrm{FUV}$ for all the regions. This is expected as NUV traces low-mass stars which are more common in galaxies \citep{gil2007}. Seyfert 2 AGN can contribute to $\sim$1-10\% of UV flux \citep{colina1997}. Hence, the SFR calculated for the whole galaxies i.e regions 1 and 2 (S1 and S2 respectively) are the upper limits as the UV emission from AGN (if present, see section \ref{discussion}) can be present. Further, the SFR of regions 3 and 4 are higher than the SFR from other selected regions. The average uncertainty in the SFR calculation is less than 5\%. Foreground extinction correction was carried out assuming a local Milky Way-like extinction curve \citep{calzetti2000}, where the A$_{UV}/$E(B-V) estimate is from \citet{bianchi2011}.

\begin{table*}
\centering
\caption{{\bf SFR calculation from UVIT obs:} Column 1: Visually selected uv bright region number corresponding to Figure \ref{nuv}; Column 2: NUV count per second (CPS); Column 3: Corresponding NUV flux; Column 4: NUV luminosity; Column 5: Star-formation rate (SFR) from NUV luminosity; Column 6: FUV count per second (CPS); Column 7: Corresponding FUV flux; Column 8: FUV luminosity; Column 9: Star-formation rate (SFR) from FUV luminosity.}
\label{tab_uvsfr}
\resizebox{\textwidth}{!}{
\begin{tabular}{ccccccccc}
\hline \begin{tabular}[c]{@{}c@{}} Region \\ No\end{tabular}   &\begin{tabular}[c]{@{}c@{}} NUV \\CPS\end{tabular}  & \begin{tabular}[c]{@{}c@{}}Flux$_{NUV}$ \\(erg/s/cm$^2$/\AA) \end{tabular} & \begin{tabular}[c]{@{}c@{}}L$_{NUV}$ \\(erg/s) \end{tabular} & \begin{tabular}[c]{@{}c@{}}SFR$_{NUV}$ \\ ($10^{-3}M_\odot yr^{-1}$)\end{tabular} &\begin{tabular}[c]{@{}c@{}} FUV \\ CPS ($10^{-1}$)\end{tabular}  & \begin{tabular}[c]{@{}c@{}}Flux$_{FUV}$\\(erg/s/cm$^2$/\AA)\end{tabular} 
& \begin{tabular}[c]{@{}c@{}}L$_{FUV}$ \\(erg/s)\end{tabular} & \begin{tabular}[c]{@{}c@{}}SFR$_{FUV}$\\ ($10^{-3}M_\odot yr^{-1}$) \end{tabular} \\ \hline
%& SFR$_{FUV}$ \hline
1	&	   4.52 & 1.09e-15 & 9.84e+41 & 95.87 & 2.98   & 1.44e-15 & 7.42e+41 & 33.63\\
2   &	   1.65 & 4.00e-16 & 3.60e+41 & 35.06 & 1.54   & 7.44e-16 & 3.38e+41 & 17.37\\
3   &	   0.43 & 1.05e-16 & 9.49e+40 & 9.24  & 0.30   & 1.46e-16 & 7.52e+40 & 3.40\\
4 	&	   0.31 & 7.59e-17 & 6.82e+40 & 6.64  & 0.22   & 1.10e-16 & 5.68e+40 & 2.57\\
5   &	   0.19 & 4.59e-17 & 4.13e+40 & 4.02  & 0.14   & 7.14e-17 & 3.68e+40 & 1.66\\
6	&	   0.13 & 3.17e-17 & 2.85e+40 & 2.78  & 0.14   & 7.00e-17 & 3.60e+40 & 1.63\\
7   &      0.08 & 1.90e-17 & 1.71e+40 & 1.67  & 0.10   & 5.18e-17 & 2.66e+40 & 1.20\\
8   &      0.09 & 2.17e-17 & 1.95e+40 & 1.90  & 0.13   & 6.65e-17 & 3.42e+40 & 1.55\\
9   &      0.09 & 2.21e-17 & 1.99e+40 & 1.94  & 0.32   & 6.60e-17 & 3.10e+40 & 1.40\\
10  &      0.05 & 1.23e-17 & 1.11e+40 & 1.08  & 0.06   & 3.14e-17 & 1.61e+40 &  0.73\\ \hline
\end{tabular}}
\end{table*}

\subsection{Star formation rates from Radio, H{$\alpha$} and IR} \label{sfr}
In this section, we have calculated the SFR from the H$\alpha$, IR, and the radio emission of the sources. These SFR values, along with UV SFR (section \ref{uv}) can help us determine the origin of the nuclear emission. 
We have  obtained the $H_\alpha$ fluxes from the HCT spectra which is given in Table \ref{tab_optical}. The SFR relation from \citet{Kennicutt.etal.1998} is:
\begin{equation}
\mathrm{SFR}_\mathrm{H\alpha} (\mathrm{M}_{\sun}~\mathrm{yr}^{-1}) = 7.9\times10^{-42}~\mathrm{L}_\mathrm{[H\alpha]}~(\mathrm{erg~s^{-1}}) \nonumber
\end{equation}

The H$_\alpha$ SFR turns out to be 0.13 M$_{\odot}$~yr$^{-1}$ and 0.30 M$_{\odot}$~yr$^{-1}$ for the nuclear regions of S1 and S2 respectively.

Next, we have calculated the SFR from mid-IR emission using the Wide-field Infrared Survey Explorer (WISE) observation \citep{wright2010}. The following relation is used from \citep{calzetti2013} 
\begin{equation}
\mathrm{SFR}_\mathrm{IR} (\mathrm{M}_{\sun}~\mathrm{yr}^{-1}) = 1.27\times 10^{-38}[\mathrm{L}_{24 \mu m} (\mathrm{erg~s^{-1}})]^{0.8850} \nonumber
\end{equation}

The 24 $\mu$m  or W4 band magnitude is taken from Infrared Science Archive (IRSA\footnote{https://irsa.ipac.caltech.edu/Missions/wise.html}). We have converted the magnitude to flux using the zero magnitude fluxes from \citet{jarrett2011}. The W4 band magnitudes are 4.53 and 5.29 which leads to the flux values of $7.84\times10^{-20}$ W/cm$^2$/$\mu m$ and  $3.87\times10^{-20}$ W/cm$^2$/$\mu m$ for S1 and S2 respectively. Hence, the calculated SFRs from the WISE W4 band for S1 and S2 are 0.58 and 0.32 M$_\odot$~yr$^{-1}$ respectively.

We have followed the \citet{condon1992} relation to calculate the SFR from the radio flux density:
\begin{equation}
\mathrm{SFR}_\mathrm{Radio} (\mathrm{M}_{\sun}~\mathrm{yr}^{-1}) = \mathrm{L(W~ Hz^{-1}})/5.3\times10^{21} \nu(\mathrm{GHz})^\alpha \nonumber
 \end{equation}
We have used the radio flux density from the uGMRT map at 1.4 GHz (Table \ref{tab_rad}), assuming $\alpha=-0.8$ to calculate the SFRs of S1 and S2, as the sources are unresolved in the FIRST image. The SFR$_{radio}$ are 0.98 M$_{\odot}$~yr$^{-1}$ and 0.65 M$_{\odot}$~yr$^{-1}$ for S1 and S2 respectively. 

The SFRs from H$\alpha$ is higher than UV which indicate that both the galaxies have recent massive star-formation which can result from the ongoing merger. If the dust is heated by the young stars and emits in IR, the SFR$_{\mathrm{H}\alpha}$ and SFR$_\mathrm{IR}$ should agree with each other \citep{dopita2003}. We find that the SFR$_\mathrm{H\alpha}$ and SFR$_{\mathrm{IR}}$ to be similar for S2 galaxy. The higher SFR$_{IR}$ in S1 also indicate the dust obscuration. However, the effect of internal dust extinction which has not been accounted for and often difficult to estimate and needs an elaborate SED modeling \citep{calzetti2013}.
%The excess SFR$_{radio}$ from other indicators can be due to the presence of AGN activity. However, these small discrepancies cannot confirm that.}

\subsection{The SMBH masses}\label{smbh_mass}
There are different methods for calculating the SMBH masses in galaxies. One commonly used method is the SMBH mass estimation from the width of the broad line component of the H$\alpha$ emission line. However, in the case Mrk 212, the broad line component of H$\alpha$ is absent for both S1 and S2. However, it is possible to determine an upper limit to the SMBH masses of S1 and S2 using the relation between SMBH mass and bulge velocity dispersion ($\sigma_\star$) \citep{mcConnell2013}. The resolution of the HCT spectrum is not good enough for the determination of $\sigma_{\star}$ for the two bulges. For S2, there is SDSS spectrum available but not for S1.
The $\sigma_\star$ of S2 from the SDSS spectrum has a value of $\sigma_\star$ = 127.6$\pm$4.04~km s$^{-1}$. This provides an upper limit to the mass of the SMBH in S2 which is $(1.62\pm0.33)\times10^{7}$ M$_{\sun}$.
Recently, SDSS-IV MaNGA\footnote{https://www.sdss.org/dr16/manga/manga-data/} DR16 has released the data for S1 only. We have obtained a stellar velocity dispersion ($\sigma_{\star}$) of 101.2$\pm$2.8~km~s$^{-1}$ and calculated the mass of the SMBH to be  $(6.7\pm1.0)\times10^6$ M$_{\sun}$. Here, the value of $\sigma_{\star}$ is derived over central 3.5$^{\prime\prime}$ $\times$ 3.5$^{\prime\prime}$ region of the source S1 using MaNGA stellar velocity dispersion map provided by the Data Analysis Pipeline \citep[DAP;][]{Westfall2019}.

\section{Discussion}\label{discussion}
\subsection{The Nature of the Dual Nuclei in Mrk 212}
There is strong evidence that AGN activities are correlated with mergers \citep{fu2018}. Hence, several studies are following up galaxy pairs where a merger is ongoing. Some recent studies have used large samples of galaxy pairs to estimate the relative fraction of AGN in isolated and merging galaxies. \citet{koss2012} have found 5\% of their sample (z$\sim$0.4) are DAGN at separations $<15$ kpc. \citet{fu2018} have found that $\sim$24\% of the paired galaxies are AGN while $\sim$13\% are binary AGNs. However, dust obscuration may be important while estimating AGN and SF pairs in mergers. \citet{satyapal2017} have found that the DAGN have higher degrees of obscuration. In contrast, \citet{gross2019} have found an opposite result where the enhanced obscuration in galaxy mergers is not due to large-scale gas inflows. 

The SDSS optical and FIRST radio images show that Mrk 212 is merger remnant with two nuclei that are bright at optical and radio wavelengths. In this subsection, we discuss whether our multi-wavelength observations of Mrk 212 indicates that it is a DAGN galaxy or not.

\subsubsection{S1}
The 15 GHz image shows an extended double-lobed structure of size $\sim$750 parsec associated with S1 (Figure~\ref{ku_pan}). Its average spectral index is steep and has the value $\alpha_{8.5}^{1.4}$= $-0.81\pm0.10$, which is consistent with optically thin synchrotron emission \citep{condon1992}. These characteristics resemble a compact symmetric object \citep[CSO;][]{perucho2002}. However, the radio power ($\sim10^{21}$~W~Hz$^{-1}$) is lower than the radio power expected from CSOs \citep{odea1998}. The radio power of S1 falls in the range of the low-luminosity AGN and its radio morphology and extent closely resembles the well-known Seyfert galaxy, NGC~1068 \citep{gallimore1996}. \citet{jarvis2019} have found their radio-quiet quasar sample to have sizes and radio powers intermediate between radio-loud AGN \citep{an2012} and low-luminosity AGN \citep{gallimore2006}. 
\citet{nyland2016} have observed a sample of early type galaxies at 5~GHz and used optical emission line diagnostics to understand the origin of nuclear emission. While many of the LINER-AGN in that sample show extended radio structures, radio cores are detected in many AGN+SF nuclei, supporting the presence of an AGN. Finally, the 1.5~GHz eMERLIN survey of nearby galaxies carried out recently by \citet{baldi2018} has detected jetted ``HII galaxies'', which have SF-dominated nuclei based on optical emission lines. However, these galaxies fall in the region occupied by LINERS in the fundamental plane (FP) of BH activity. Our results are consistent with the findings of \citet{baldi2018}. 

At high frequencies, most of the non thermal emission from SF galaxies are expected to be resolved out with only thermal emission being detected on hundreds of parsecs scale \citep{murphy2018}. The average power of individual radio supernova is $2\times10^{20}$~W~Hz$^{-1}$ \citep{bondi2005}. Hence, it is unrealistic to have a multiple supernovae producing a radio luminosity of $\sim10^{21}$~W~Hz$^{-1}$ and a two-sided radio structure of ~750 pc centred on the optical centre. We conclude that the radio source associated with S1 is a CSO-like weak AGN. 

The HCT optical spectrum of S1 lies in the composite (AGN + SF) region of the BPT diagram. Although the BPT diagram is an essential tool to confirm the presence of an AGN, several studies have shown that AGN can be obscured by dust and gas accretion during mergers. For example, \citet{satyapal2017} have found that their mid-IR selected DAGN sample do not show AGN signatures in their optical spectra. A fraction of X-ray detected AGN are also found in the star-forming (HII) region of the BPT diagram \citep{agostino2019}. We have compared the SFR calculated using different tracers -- H$\alpha$, UV, and radio. The SFR from radio data is $\sim$0.98 M$_{\odot}$~yr$^{-1}$ while the  H$\alpha$, NUV, FUV and IR SFRs are $\sim$0.13, 0.095, 0.033 and 0.58 M$_{\odot}$~yr$^{-1}$ respectively. We believe that the higher SFR$_\mathrm{Radio}$ indicates the presence of a weak AGN in S1. Its soft X-ray spectrum may indicate the presence of absorption by gas and dust.

\subsubsection{S2}
A compact core is detected in the 15~GHz VLA image (C in Figure \ref{x_band}) which coincides with the optical centre and nucleus S2. Its radio luminosity is $\sim10^{21}$~W~Hz$^{-1}$, qualifying it to be a weak AGN. The relative prominence of this component at the highest observed frequency (15~GHz) indicates that it has an inverted spectrum, as expected from the synchrotron self-absorbed base of an unresolved jet. 

We find that the optical HCT spectrum puts S2 in the composite (AGN + SF) region. However, \citet{hern2016} have also fitted the SDSS spectra of S2, calculated the flux ratios and found that S2 lies in the AGN region of the BPT diagram. Based on the detection of the compact radio core as well as optical spectra, we conclude that S2 is a weak AGN.

The presence of DAGN in Mrk 212 is supported by high-resolution radio images and optical spectroscopy. In Mrk 739, one of the nuclei of DAGN did not show any evidence of AGN activity in optical or radio wavelengths. However, its X-ray emission was explained by an AGN \citep{wong2016}. Therefore, these studies clearly show that discarding the presence of a DAGN after following up DAGN candidates in single waveband observations, can lead to a low-detection rate of DAGN.

\subsection{The Extended Radio Structure around S2}
An extended radio structure is detected in the 8.5~GHz VLA image (components A, B in Figure \ref{x_band}), although its presence is also hinted at in the uv-tapered 15~GHz image. The spectral index for components A and B close to S2 is $\alpha=-0.53\pm0.10$ and $-0.34\pm0.10$, respectively. These relatively flat spectral indices are consistent with emission from SF regions where free-free emission could be mixed with synchrotron emission from supernovae.

Models of star-formation predict an increase in the SFR in the disks, nuclei and in the outer tidal tails \citep{duc2000}. We have calculated the SFR of visually selected regions of Mrk~212 using the UVIT data (Table \ref{tab_uvsfr}). As mentioned in section \ref{uv}, the regions 3 and 4 which are situated close to S2 nucleus, have higher SFR compare to other regions. These two regions coincide with the extended radio components A, B (Figure \ref{x_band}) detected at 8.5 GHz. We have compared the SFRs of components A and B derived from radio and UV emission. The 1.4~GHz uGMRT flux densities of A and B are 0.53 mJy and 0.37 mJy respectively. The corresponding SFRs are 0.13 M$_{\odot}$~yr$^{-1}$ and 0.06 M$_{\odot}$~yr$^{-1}$, respectively, which are 10-14 times higher than the UV estimates. However, with present data, it is not clear that there is any connection between the UV blobs and AGN activity in S2, or if they are a result of the merger. Future observations of molecular gas distribution in this system can help us to understand star-formation. A more sensitive high resolution radio study can reveal a connection, if present, between the structures detected at 15 GHz and 8.4 GHz.

\section{Summary and conclusions}
We have carried out a multi-wavelength study of the merging galaxy Mrk~212 that possesses two optical nuclei S1 and S2 at a projected separation of $\sim11.8\arcsec$ or $\sim6$~kpc. Our main findings are the following:
\begin{enumerate}[label=(\arabic*)]
\item The 15 GHz VLA image of Mrk~212 reveals a double radio source associated with S1 and a compact radio structure (C) associated with S2. An extended radio structure is detected at 8.5~GHz that is located 1$\arcsec$ away from S2. 

\item Using HCT optical observations, we find that S1 and S2 both fall in the AGN+SF region, in the BPT diagram. The Chandra X-ray data reveal a soft spectrum for S1 and a harder spectrum for S2. 

\item Using deep UVIT FUV and NUV observations, we were able to resolve the star-forming knots in S2 and in the tidal tails of Mrk~212. The SF knots in S2 coincides with the two sided radio structure (A, B) detected at 8.5 GHz. The SFR was calculated for the UV knots, using H$\alpha$, IR and the radio flux densities. The IR and radio estimates are broadly consistent with each other.

\item The double radio source associated with S1 at 15 GHz with the $1.4-8.5$~GHz spectral index ($-0.8\pm0.10$) is consistent with a CSO-like weak AGN. The presence of the compact radio core and the presence of AGN emission lines in the optical spectrum of S2, support the presence of another AGN in S2.

\item The extended radio structure observed in the 8.5~GHz image, offset from the optical nucleus S2, has a relatively flat spectral index. These regions coincide with UV star-forming knots. Therefore, we conclude that these represent star-forming regions.

\end{enumerate}

\section{ACKNOWLEDGEMENTS}
We thank the referees for constructive and insightful suggestions that have significantly improved the quality of the manuscript. We acknowledge IIA and NCRA for providing the computational facilities. The National Radio Astronomy Observatory is a facility
of the National Science Foundation operated under cooperative
agreement by Associated Universities, Inc. We thank the staff of the GMRT  who have  made  these observations possible. GMRT is run by the National Centre for Radio Astrophysics of the Tata Institute of Fundamental Research. The optical observations were done at the Indian Optical  Observatory (IAO) at Hanle. We thank the staff of IAO, Hanle and CREST, Hosakote, that made these observations possible. The facilities at IAO and CREST are operated by the Indian Institute of  Astrophysics, Bangalore. This publication uses the data from the AstroSat mission of the Indian Space Research Organisation (ISRO), archived at the Indian Space Science DataCentre  (ISSDC)  which is a result of collaboration between IIA, Bengaluru, IUCAA, Pune, TIFR, Mumbai,several  centres of ISRO, and CSA.  This project makes use of the MaNGA-Pipe3D dataproducts. We thank the IA-UNAM MaNGA team for creating this catalogue, and the Conacyt Project CB-285080 for supporting them. Operation of the Pan-STARRS1 telescope is supported by the National Aeronautics and Space Administration under Grant No. NNX12AR65G and Grant No. NNX14AM74G issued through the NEO Observation Program. This research has made use of the NASA/IPAC Extragalactic Database (NED),
which is operated by the Jet Propulsion Laboratory, California
Institute of Technology, under contract with the National Aeronautics and Space Administration. Funding for the Sloan Digital Sky Survey IV has been provided by
the Alfred P. Sloan Foundation, the U.S. Department of Energy Office of Science,
and the Participating Institutions. SDSS- IV acknowledges support and resources 
from the Center for High-Performance Computing at the University of Utah. 
The SDSS web site is www.sdss.org. We acknowledge the support of the Department of Atomic Energy,Government of India, under the project 12-R\&D-TFR-5.02-0700.

\section{DATA AVAILABITY}
The data underlying this article will be shared on reasonable requestto the corresponding author. The raw data can be obtained from the following data archive: (i) https://archive.nrao.edu/archive/advquery.jsp
(ii) https://naps.ncra.tifr.res.in/goa/data/search (iii) https://astrobrowse.issdc.gov.in/astro$_a$rchive/archive/Home.jsp

\appendix
\section{Appendix}
Here we present additional radio images of Mrk~212 at multiple radio frequencies with the VLA and uGMRT.

\begin{figure*}
\includegraphics[width=10cm,trim=10 30 10 0]{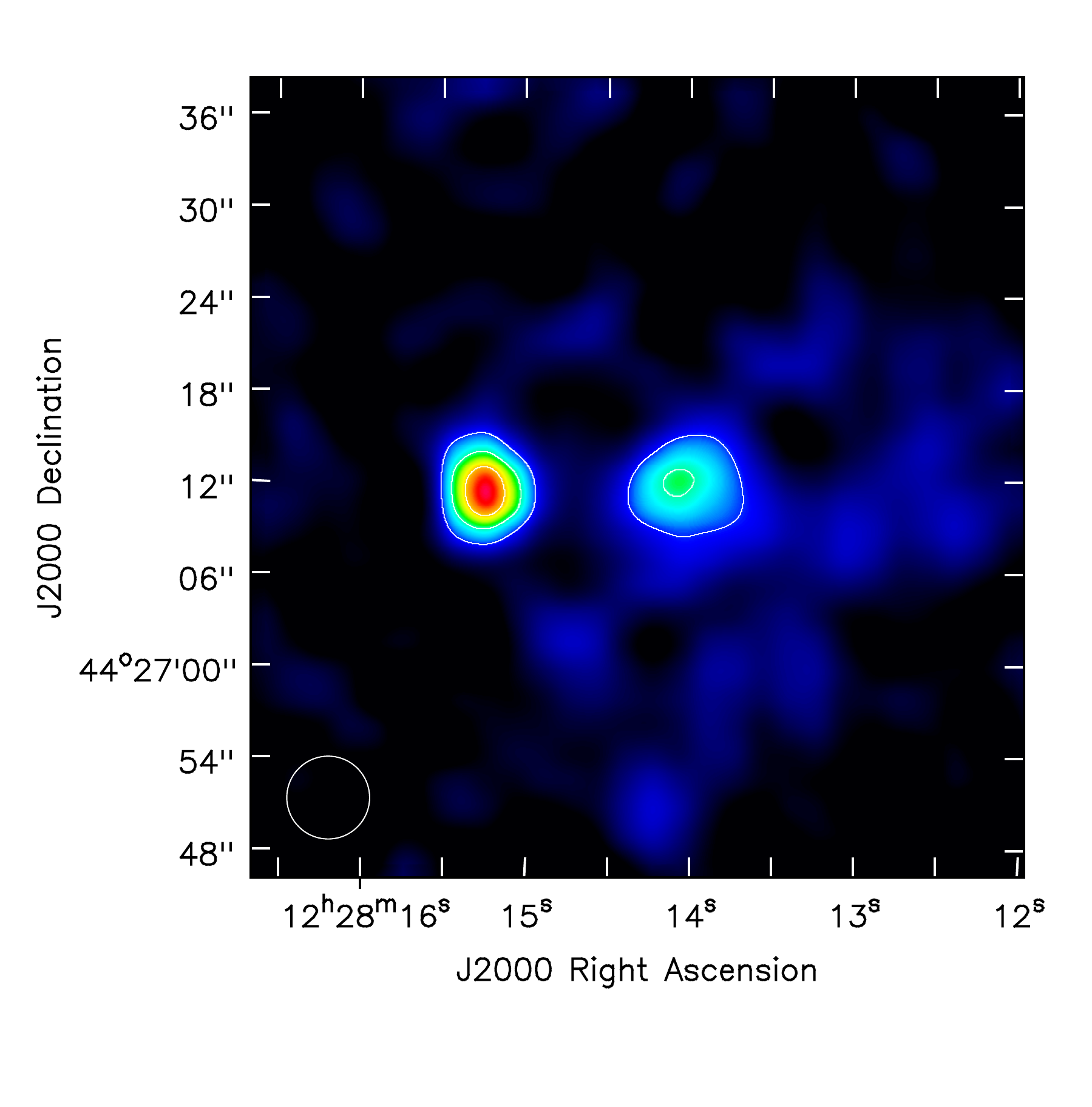}
\caption{VLA FIRST 1.4 GHz image of Mrk~212 with radio contours. The contours are 20, 40, 60, 80 \% of the peak flux density value 2.15~mJy~beam$^{-1}$.}
\label{first}
\end{figure*}

\begin{figure*}
\includegraphics[width=10cm,trim=10 30 10 0]{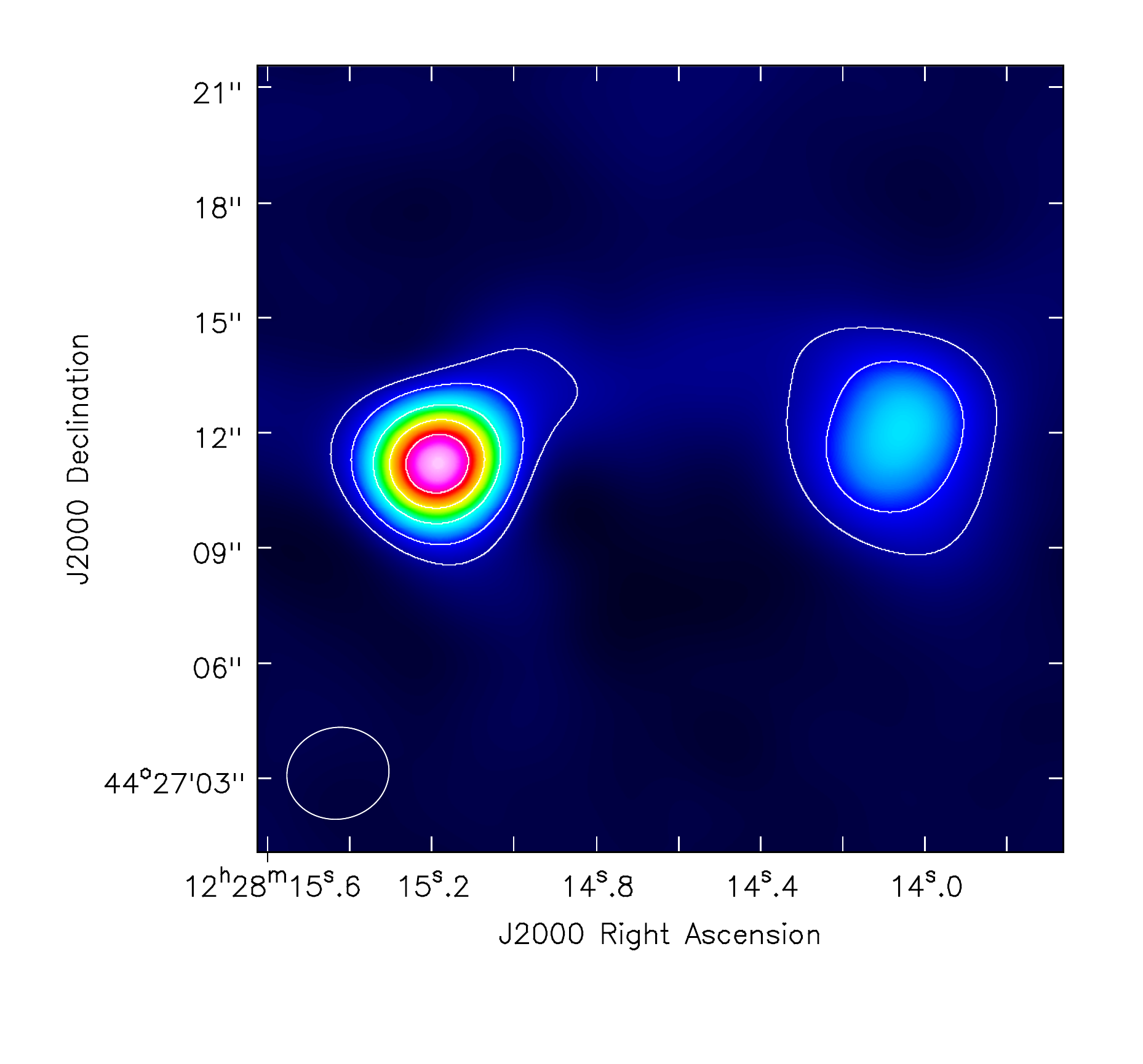}
\caption{uGMRT 1.4~GHz image with contours. Both the sources are detected in this 1.4 GHz image. The contours are 20, 40, 60, 80\% of the peak flux density 2.68~mJy~beam$^{-1}$.}
\label{gmrt}
\end{figure*}

\begin{figure*}
\includegraphics[width=8cm]{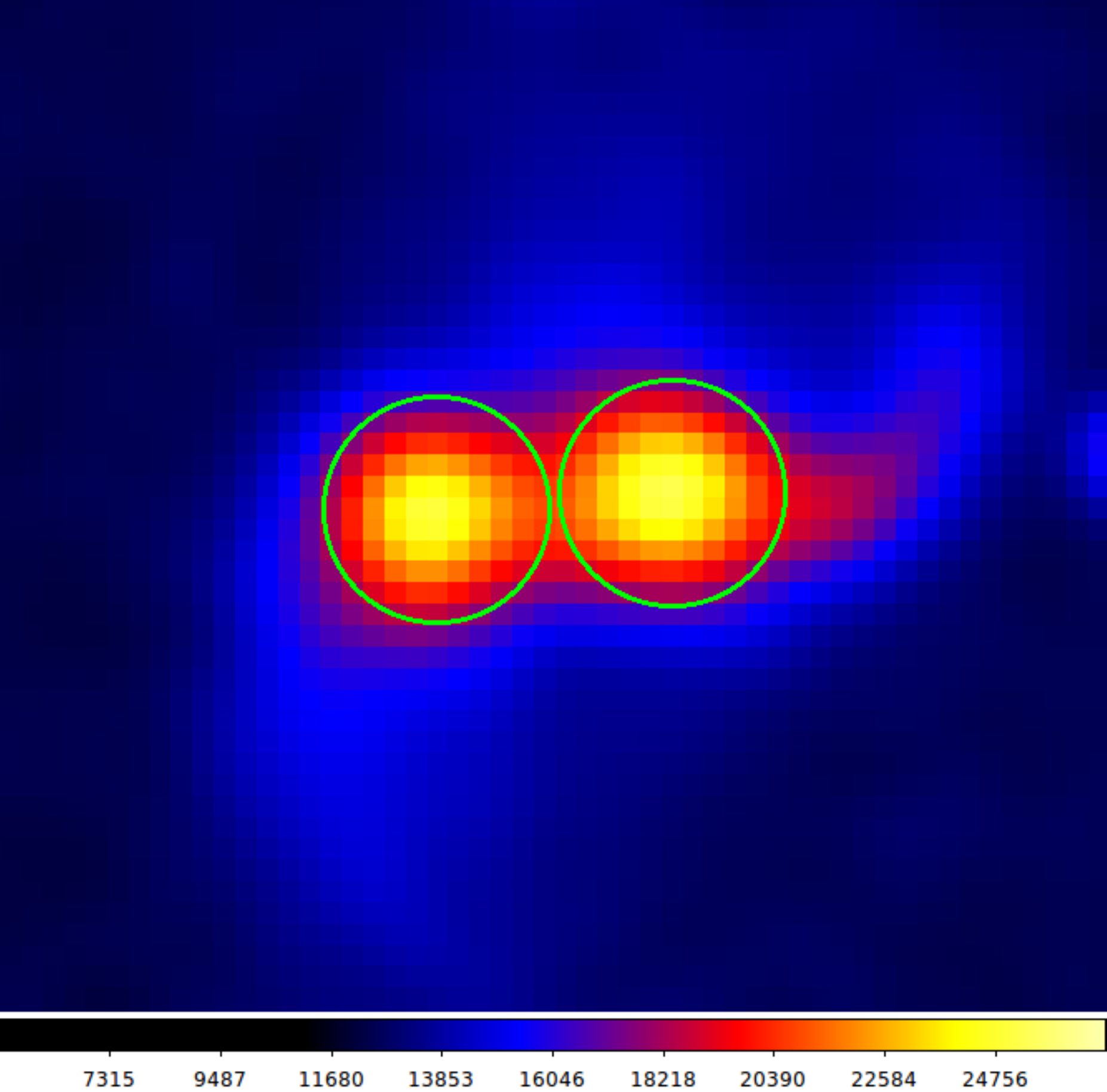}
\caption{The optical DSS image of Mrk~212. The two regions marked by green circles were used to measure the hardness ratio of each core in the X-ray analysis.}
\label{dss_regions}
\end{figure*}

%\bibliographystyle{mn2e}
 %\bibliography{MRK212}
\end{document}